# Vehicle Noise: Comparison of Loudness Ratings in the Field and the Laboratory

Gerard Llorach[†,1,2,3], Dirk Oetting[1,2], Matthias Vormann[1,2], Markus Meis[1,2], and Volker Hohmann[1,2,3]

{gerard.llorach.to@uni-oldenburg.de, oetting@hz-ol.de, vormann@hz-ol.de, meis@hz-ol.de, volker.hohmann@uni-oldenburg.de}

[1] Hörzentrum Oldenburg gGmbH
Oldenburg, Germany

[2] Cluster of Excellence Hearing4All
Dept. of Medical Physics and Acoustics
University of Oldenburg
Oldenburg, Germany

[3] Auditory Signal Processing
Dept. of Medical Physics and Acoustics
University of Oldenburg
Oldenburg, Germany

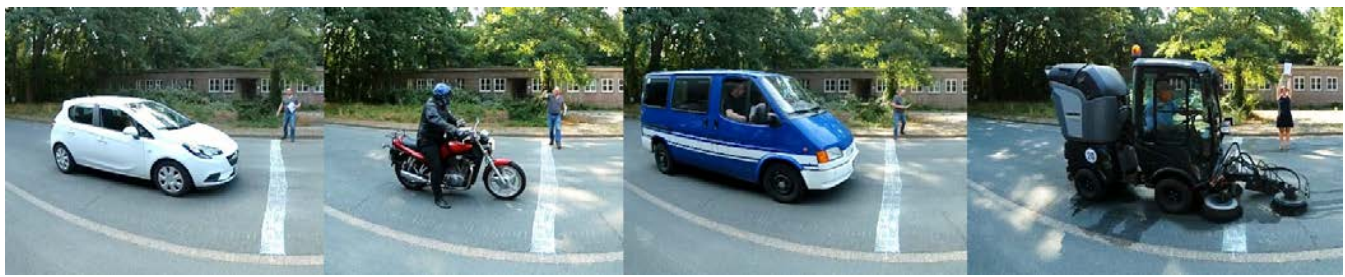

**Figure 1. Vehicles used in the experiment. From left to right: car (Opel Corsa 2016), motorbike (Suzuki VX 800 800cc 1994), van (Ford Transit FT100 1999), and street sweeper (Kärcher MC 50). The figure is taken from [7].**

## ABSTRACT

**Objective:** Distorted loudness perception is one of the main complaints of hearing aid users. Being able to measure loudness perception correctly in the clinic is essential for fitting hearing aids. For this, experiments in the clinic should be able to reflect and capture loudness perception as in everyday-life situations. Little research has been done comparing loudness perception in the field and in the laboratory.

**Design:** Participants rated the loudness in the field and in the laboratory of 36 driving actions done by four different vehicles. The field measurements were done in a restricted street and recorded with a 360º camera and a tetrahedral microphone. The recorded stimuli, which are openly accessible, were presented in three different conditions in the laboratory: 360º video recordings with a head-mounted display, video recordings with a desktop monitor, and audio-only.

**Sample:** Thirteen normal-hearing participants and 18 hearing-impaired participants participated in the study.

**Results:** The driving actions were rated significantly louder in the laboratory than in the field for the audio-only condition. These loudness rating differences were bigger for louder sounds in two laboratory conditions, i.e., the higher the sound level of a driving action was the more likely it was to be rated louder in the laboratory. There were no significant differences in the loudness ratings between the three laboratory conditions and between groups.

**Conclusions:** The results of this experiment further remark the importance of increasing the realism and immersion when measuring loudness in the clinic.

## 1 Introduction

One of the common complaints of hearing-impaired listeners is about loudness: some sounds are too loud and others are not heard [1]. When the listeners are rehabilitated with hearing aids, the hearing devices are fitted and adjusted in the clinic with controlled acoustic situations and audiometric tests, which are far from reflecting real-life scenarios. These disparities, between the clinic and the field, may lead to inaccurate estimates of loudness perception and, in consequence, to inappropriate settings in the hearing aids [2].

To overcome these problems, loudness-related measurements in the laboratory should become more ecologically valid [3] than established methods, i.e., they should better reflect real-life loudness perception. Loudness perception differences between the field and the laboratory have been rarely studied, as the complexity of a field situation is rather difficult to reproduce in the laboratory. Among the few existing studies, the experiment from [4] showed some interesting disparities between the field and the laboratory. Normal-hearing listeners and listeners with hearing loss were instructed to use research hearing aids in the field for a week. They could adjust the loudness through volume control, and, when they did, the research hearing aid recorded the volume gain of the device and the sound pressure level of the field situation. Following, the participants were invited to the laboratory, where they had to adjust the volume of their research hearing aids, this time in a controlled audiovisual laboratory experiment. The stimuli in the laboratory, which consisted of recordings of a bushwalk, an office situation, a small gathering, a motorway, and sawing wood with a power tool, were presented through a television screen and two frontal loudspeakers. The normal-hearing participants chose lower gains in the laboratory than in the field, whereas the listeners with hearing loss did the opposite: they chose higher gains in the laboratory than in the field. Several explanations are given in the article, such as the difficulty of imagining to be in a particular situation in the laboratory, the possibility of the listeners with hearing-loss using lower gains in the field because of undesired soft background noises, and the possibility of the normal-hearing listeners using higher gains in the field to compensate for the reduced frequency range of the hearing aids. These findings raised several questions about loudness perception differences in the field and the laboratory.

An important factor of measuring loudness perception in the laboratory is visual information: visual cues have been found to influence loudness perception. When sounds were presented together with congruent visual cues, they were usually perceived as less loud [5]. In further experiments, the differences between immersive audiovisual simulations (i.e., a car simulator and videos via a head-mounted display) and audio-only reproduction were investigated. The loudness judgments were decreased by about 15% in the immersive audiovisual simulations, and, in some individual cases, by more than 50%. These findings were further confirmed in similar experiments, reviewed in [6].

The aim of our work is to measure the differences in loudness perception between the field and different laboratory setups, and to further explore the different factors influencing loudness perception in laboratory experiments. Doing precise loudness measurements in the field is more complicated than in the laboratory, as it is more difficult to control and reproduce systematically the same stimuli. Nevertheless, an experiment in the field is closer to a real-life situation and may better represent the perception outside the laboratory. We measured loudness perception in the field and in the laboratory with the same participants. We recorded the stimuli in the field and replicated it in the laboratory with different setups. The laboratory setups ranged from immersive experiences (head-mounted display and stereo audio) to more simplistic clinical setups (only audio with a single frontal loudspeaker), as we wanted to know which requirements a clinical setup should have to measure loudness perception as in the field.

The methods and results of the field experiment have already been published and can be found in [7] for the normal-hearing listeners and in [8] for the listeners with hearing loss. Our work provides an addition to the findings of [4], where a direct comparison between the stimuli in the laboratory and the field could not be done, due to the uncontrolled nature of the field situations; and to the work of [9], where there were no field measurements to compare to the audiovisual simulations. To the best of our knowledge, this is the first work that compares field and laboratory loudness perception using the same stimuli and the same participants. The implications for the fitting procedures for the listeners with hearing loss will be not discussed in this paper.

## 2  Method

The participants were asked to rate the perceived loudness of different driving actions, using the response scale of the categorical loudness scaling (CLS) procedure [10] for loudness. The CLS consists of an ordinal scale with name tags from "not heard" and "very soft", to "loud" and "extremely loud". The field experiment was conducted in a reserved street on a former military facility. The participants were distributed in four different sessions / dates. The listening positions were on a side of the street, and the participants rated the driving actions of four different vehicles (see Figures 1 and 2). These driving actions were recorded with a 360º camera (Xiaomi Mi Sphere Camera, Xiaomi, Hong Kong), a tetrahedral microphone (Core Sound TetraMic, Core Sound, LLC, Teaneck, USA), and a sound level meter.

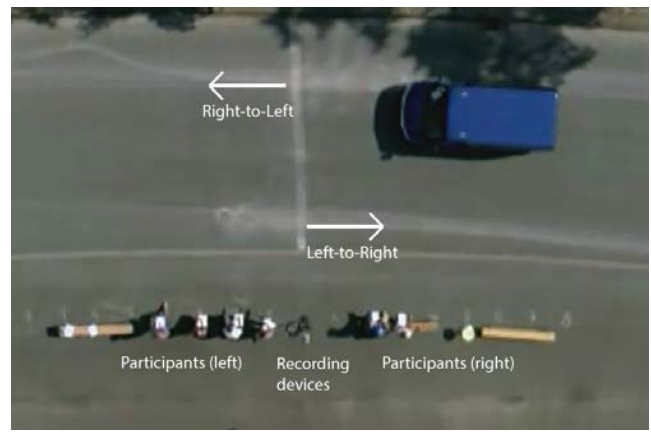

**Figure 2. Setup of the field experiment. The figure is taken from [7].**

In the laboratory experiments, the recorded driving actions were played back in three different conditions: (1) 360º video playback with a head-mounted display and stereo audio (360VID); (2) video playback with a computer monitor and stereo audio (2DVID); and (3) audio-only with a frontal loudspeaker (AO). These conditions are summarized in Table 1.

**Table 1. Conditions tested in the laboratory experiment.**

| Condition abbreviation | Visual setup | Audio setup |
|---|---|---|
| 360VID | Head-mounted display with 360º videos | Stereo with loudspeakers at +- 60º |
| 2DVID | Computer monitor with 2D videos | Stereo with loudspeakers at +- 60º |
| AO | None | Mono (0º) |

With such design it is not possible to discern the effect of visual cues independently, as the audio setup is different in the audio-only condition. Rather than measuring the effect of visual cues, the design of this experiment compares two audiovisual setups and a setup (AO) that represents the most simplistic clinical setup for loudness measurements. Because the audiovisual setups have the same audio setup, a comparison between the visual display (HMD and computer monitor) is possible.

## 2.1 Participants

Thirteen normal-hearing listeners (six female and seven male) and 18 listeners with hearing loss (11 female and seven male) participated in the field and in the laboratory experiments. The normal-hearing participants had a pure-tone average at the four frequencies 500, 1000, 2000, and 4000 Hz between -2 and 13 dB HL. The pure-tone average of the hearing-impaired listeners was between 34 and 52 dB HL with an average of 42.36 dB HL. The difference between the pure-tone average of the left and right ear was below 15 dB so all listeners had symmetric hearing loss. Hearing aids were fitted with trueLOUDNESS (program 1) and with NAL-NL2 (program 2) [11]. In the laboratory experiments, the trueLOUDNESS fitting was used. Details of the hearing-aid fitting and a description of the hearing-impaired listeners can be found in [8]. Ethical approval of the experiment was granted by the ethics committee of the University of Oldenburg. The participants were recruited, contacted, and reimbursed through Hörzentrum GmbH.

## 2.2 Stimuli

Four different vehicles were used, which are shown in Figure 1: a white car (Opel Corsa 2016), a red motorbike (Suzuki VX 800 800cc 1994), a dark blue van (Ford Transit FT100 1999), and a street sweeper (Kärcher MC 50). Loudness for the first three vehicles (the car, the motorbike, and the van) were rated in 10 different conditions (five different driving actions, once on each side of the street). These actions were "stand by with the engine on", "stand by to drive forward", "pass by at 30 km/h", "pass by at 50 km/h", and "break until stopping". The vehicles were driving towards the end of the street and turning back, once out of the sight of the participants, to do the next driving action, this time on the other side of the street. For example, a vehicle would "stand by to drive forward" on the participant's street side, reach the end of the street, turn back, and "pass by at 30 km/h" on the other side of the street. Loudness ratings for the street sweeper were assessed in six driving actions (three different actions, once on each side of the street): "stand by with the engine on", "stand by with the brushes on", and "stand by to move and brush forward".

**Table 2. Vehicle's driving actions with average maximum dB SPL (125ms windows). The actions are numbered with the order of presentation during the experiment. LR and RL stand for the direction of the driving: Left-to-Right (LR) and Right-to-Left (RL). The table is taken from [7].**

| | dB SPL of the driving actions in the field | | | | | | | | | |
|---|---|---|---|---|---|---|---|---|---|---|
| | 1A. Stand by (close) | 2A. Accelerate LR (close) | 3A. 30 km/h RL (far) | 4A. 50 km/h LR (close) | 5A. Break and stop RL (far) | 6A. Stand by (far) | 7A. Accelerate RL (far) | 8A. 30 km/h LR (close) | 9A. 50 km/h RL (far) | 10A. Break and stop LR (close) |
| **Car** | 71.2 | 84.3 | 73.3 | 81.5 | 75.2 | 67.9 | 80.1 | 75.2 | 76.9 | 77.1 |
| **Motorbike** | 83.5 | 91.5 | 82.5 | 89.7 | 81.1 | 78.4 | 86.6 | 89.0 | 88.1 | 84.0 |
| **Van** | 82.7 | 88.4 | 81.1 | 90.1 | 80.5 | 80.3 | 87.8 | 84.5 | 85.9 | 82.8 |
| | 1B. Stand by (close) | 2B. Brushes on (close) | 3B. Forward LR (close) | 4B. Stand by (far) | 5B. Brushes on (far) | 6B. Forward RL (far) | | | | |
| **Street sweeper** | 83.6 | 91.1 | 92.6 | 76.9 | 83.7 | 83.5 | | | | |

Each driving action was repeated eight times (four sessions, test and retest for the normal-hearing listeners, and program 1 and program 2 for the listeners with hearing loss). The drivers aimed at repeating the driving actions identically. The sound levels per driving action had an average standard deviation of 1.7 dB SPL and a reliability coefficient of 0.96 ($p<0.001$). The sound pressure levels of the driving actions were measured with a sound level meter (Nor140, Norsonic Tippkemper GmbH, Oelde-Stromberg, Germany) and were calculated as the maximum level in dB SPL in windows of 125ms. The average level per driving action can be seen in Table 2.

The recorded signals in the field were cut and processed for the laboratory experiment. Out of the eight recordings per driving action, the one that contained less noise and distractions (birds chirping, wind, coughing) was selected for each driving action, leading to 36 final recordings for the laboratory. Each driving action recording was edited and cut to last 12 seconds. The acoustic recordings of the Tetrahedral microphone were synthesized to a stereo format (XY microphone setup) using the VVMic software from VVAudio. The faces of the participants were blurred for anonymity in the video recordings of the 360º camera. The sound levels of the selected driving actions ranged from 67.8 to 94.6 dB SPL (maximum level in windows of 125 ms). The acoustic levels in the laboratory were adjusted using a sound level meter (Nor140, Norsonic Tippkemper GmbH, Oelde-Stromberg, Germany) to match the sound pressure levels recorded in the field. The sound level meter was placed at the approximate position of the listener's ears in the laboratory. A global gain was set for all driving actions to adjust the sound levels. Due to the room acoustics of the laboratory and the signal differences between driving actions, a variability of ±2 dBbetween the levels of the field and the laboratory was present. This sound level variability was not controlled for each driving action, as it was similar to the variability of the repetition of the driving actions (std of 1.7 dB SPL). The audiovisual recordings of the driving actions for the laboratory experiment can be found in [12].

## 2.3 Setup

In the field experiments, the participants were sitting on the side of the road where the vehicles were driving (see Figure 2). The participants were seated on benches and chairs and they kept their sitting position for the whole experiment.

In the laboratory experiment, the listeners were sitting on a chair in an acoustically treated room. They were sitting in the middle of a circle of 12 spectrally flat loudspeakers GENELEC 8030 BPM (Genelec Oy, Olvitie, Finland). The loudspeakers were at a distance of 1.2 meters from the center, at a height of 1.2 meters, and were located every 30º. In our experiment, only the loudspeakers placed in the +- 60º angles (stereo) and the frontal direction (mono) were used. For the 360VID and 2DVID conditions, the stereo loudspeakers were used. The frontal loudspeaker was used for the AO condition. In the 2DVID condition, the listeners had a computer monitor in front of them, where the videos were displayed. The computer monitor was at an approximate height of 70 cm and within the arm's reach of the listener. This computer monitor was moved away from the listener in the other two conditions because they used the head-mounted display (HMD) for the 360VID condition and they did not have any visual stimuli in the AO condition. The head-mounted display used was the HTC Vive (HTC Corporation, New Taipei City, Taiwan). The videos were reproduced with the "Media Player Classic - Home Cinema" software in the 2DVID, and with the "Steam 360 Video Player" in the 360VID condition. The computer used Windows 10 with an NVIDIA Quadro M5000 graphics card. The participants had a button on their lap that would mute the playback, in case of emergency or extreme discomfort.

## 2.4 Procedure

*2.4.1 Field experiment.* The participants were distributed in four different sessions, as there was a limited number of seats. In each session, all 36 driving actions were done, then there was a pause of 30 minutes, and the 36 driving actions were repeated. For the normal-hearing listeners, this was a test and retest of the ratings. The listeners with hearing loss were tested in the other hearing aid fitting: in the first 36 driving actions the trueLOUDNESS fitting was used, and after the pause, the NAL-NL2 fitting was used.

The participants were instructed to rate the loudness of the driving actions. A researcher indicated the number of the driving action to rate when the driving action was at its loudest instant (see video recordings in [12]).Once all participants rated the current driving action, the next driving action was executed. The driving actions followed the order shown in Table 2 and each vehicle did all its driving actions consecutively. The car started first, followed by the motorbike, the van, and the street sweeper.

*2.4.2 Laboratory experiment.* We conducted the laboratory experiments with the same participants. The field and laboratory experiments were separated by approximately 8 months. For the participants with hearing loss, the same hearing aids with the trueLOUDNESS fitting were used. An audiologist assisted with the hearing aids during the experiment.

The head-mounted display was shown and given to the participants to familiarize them with the technology. The interpupillary distance of the listeners was measured and adjusted correspondingly. The straps of the head-mounted display were adjusted to the head of the participants while the driving actions of the car were shown through the device without sound. During this adaptation phase, the participants were asked to explore the 360º environment by head movements and to make themselves comfortable with the head-mounted display. This phase lasted less than 2 minutes.

The order of the driving actions was the same as in the field experiment. After each driving action, the video was paused until the listener indicated the perceived loudness. During this pause, the driving action number and the response scale were shown in the video, and no sounds were played back. In the 360VID condition, an additional letter was added for each loudness category in the questionnaire appearing in the video. This way, the listeners could answer verbally to the questionnaire without taking off the head-mounted display.

## 2.5 Data processing

Not all participants experienced the same sound levels during the field experiment, as they were seated in different positions along the road (see Figure 2). The sound pressure levels that they experienced in the laboratory were different from the ones they were exposed to in the field for most driving actions, as the levels in the laboratory were not adjusted individually. We approximated the sound pressure level differences by assuming that the sound sources were omnidirectional and that there were no spectral differences. We used the following equation to compute the sound level differences:

$$dB_{diff} = sgn(d_2 - d_1) \cdot |20 \cdot \log\left(\frac{d_1}{d_2}\right)| \qquad (1)$$

were $dB_{diff}$ is the calculated sound level difference between the recording device and the participant, $d_1$ is the approximate distance between the position of the sound level meter and the position of the vehicle in its loudest instant of a driving action, $d_2$ is the approximate distance between the sitting position of a participant and the position of the vehicle in its loudest instant of a driving action, and sgn is the sign function, which determines if the dB difference is positive or negative. The driving actions that had equal levels for all participants (Table 1. 3A, 4A, 8A, 9A) had a 0 dB difference. The level differences between the laboratory and the field stimuli (see Equation 1) had an average of 1.9 dB with a standard deviation of 2.3 dB, with a range of -0.8 dB to 8.1 dB for all participants and driving actions.

We removed the ratings of the participants where the sound level difference was bigger than 1.5 dB. If a participant experienced a level difference above the set threshold, his/her loudness ratings of that driving action were removed in all conditions (field, HMD, 2VID, AO). Overall, 35.8 % of the ratings were removed (18.8 % NH, 17.0 % HI), with a maximum of 61.1 % for a participant. None of the 36 driving actions were completely removed. Figure 3 shows the distribution of the sound level differences.

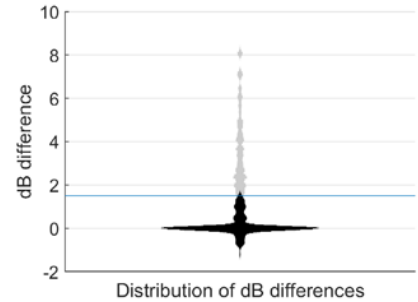

**Figure 3. Distribution of the sound level differences between the field and the laboratory for all ratings due to their sitting position and the driving actions. In black, the dB difference of the remaining ratings. In grey, the dB difference of the removed ratings. The threshold is marked with a blue horizontal line**

The remaining of the loudness ratings were averaged per participant and condition, resulting in four scores per participant. To average them, the loudness categories were transformed to a monotonically increasing numerical scale between 0 and 50 with steps of +5 for each loudness category / response alternative, as recommended by the ISO standard [10]. We assumed that the loudness categories were equidistant (see Discussion). Because the NH participants did two field measurements (test and retest), the average between the test-retest rating was used to calculate the mean rating of the field condition. For the participants with hearing loss, we averaged the field ratings that were done with the trueLOUDNESS fitting, as the same fitting was used in the laboratory conditions.

## 3 Results

In order to know if there were any differences between the field and the laboratory conditions, we performed a repeated measures ANOVA. The within-subject factor of the ANOVA was the mean of the loudness ratings per condition. The between-subject factor was the hearing type (normal hearing or hearing impaired). Mauchly's Test of Sphericity indicated that the assumption of sphericity had been violated, $\chi2(5) = 14.504$, $p = 0.013$, and therefore, a Greenhouse-Geisser correction was used. There was no interaction effect between the conditions and the hearing type, $F(2.214, 6.586) = 0.280$, $p = 0.781$. The statistical test determined that the mean loudness rating differed significantly between conditions, F(2.243, 131.618) = 5.591, p = 0.004. Pairwise comparisons with a Bonferroni correction showed that the mean loudness ratings for the field condition were significantly different from the AO condition (p = 0.018), but not from the 2DVID condition (p = 0.060) and from the HMD condition (p = 1.0). The laboratory conditions were not significantly different between them, according to the pairwise comparisons. Figure 4 shows the distribution of the mean loudness ratings of the four conditions. Overall, the loudness ratings were higher in the laboratory than in the field. The two laboratory conditions that were not significantly different from the field were the HMD and the 2DVID, which included visual cues and stereo audio. The 2DVID condition, which was less immersive than the HMD, was borderline non-significant.

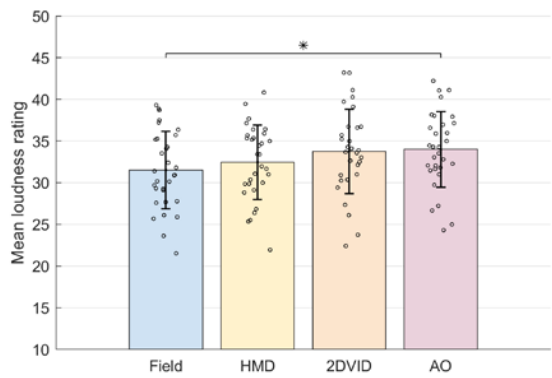

**Figure 4. Distribution of the mean loudness ratings of the participants. There are four bars representing the mean of each condition: Field, HMD, 2DVID, and AO. The vertical line in the middle of each bar indicates the standard deviation of the distribution. The black dots indicate the individual mean loudness rating for each condition. Significant differences are indicated with an asterisk (p < 0.05).**

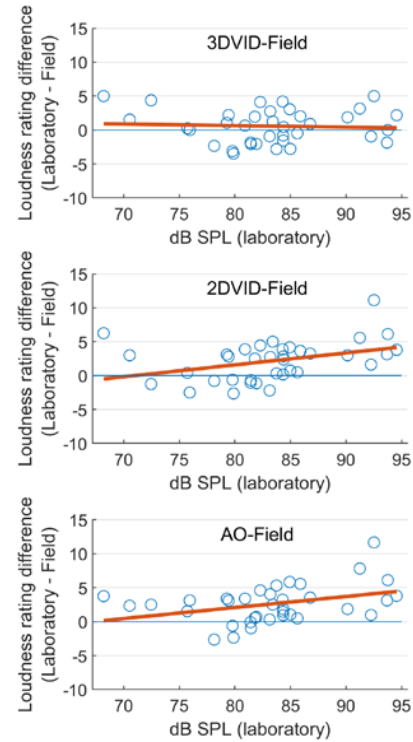

**Figure 5. Relationship between the sound levels and the laboratory-field differences in loudness ratings. The relationship with each laboratory condition is represented in a different panel: HMD (top), 2DVID (center), and AO (bottom). Each blue circle represents the mean loudness difference for a driving action. The thick orange line represents the linear relationship of the data points. If the driving action circles are above zero, these driving actions were rated louder in the laboratory.**

In order to know if these differences were more relevant for loud or soft noises, we computed the correlation between the sound pressure level of the driving actions and the laboratory-field loudness rating differences. The loudness rating differences were computed between the field and the laboratory conditions per driving action and participant. For each driving action we computed the average laboratory-field difference. Figure 5 shows the loudness laboratory-field difference for each driving action. Each circle represents the difference for a driving action. The Spearman correlation coefficient between the HMD-Field loudness rating differences and the sound pressure levels was 0.05 (p = 0.79), the 2DVID-Field loudness rating differences and the sound pressure levels was 0.43 (p < 0.01), and between the AO-Field loudness rating differences and the sound pressure levels was 0.36 (p = 0.03). If there were differences between the laboratory and the field ratings, these were higher when the sounds were louder. Hence, the difference between the laboratory ratings and the field ratings was bigger as the sound pressure level of the driving actions increased. This correlation was only found to be significant for the 2DVID-Field and the AO-Field differences.

## 4 Discussion and Conclusions

The vehicle driving actions were perceived louder in the laboratory than in the field, an effect that was only significant for the audio-only condition with a single loudspeaker. These laboratory-field differences were reduced when visual cues and stereo audio were added, thus pointing out the importance of increasing the realism and immersion in clinical evaluations. Which factor (adding visual cues or using stereo audio) had more influence in the reduction of laboratory-field loudness differences cannot be derived from this experiment. Nevertheless, according to previous research, visual cues reduce loudness perception [6], thus suggesting that visual cues had an effect in this experiment. Additionally, when comparing the two audiovisual conditions to the field ratings, the least visually immersive condition (2DVID) was borderline non-significant, suggesting that immersive visual cues further reduce field-laboratory differences.

The loudness perception differences became more apparent for higher sound levels in the AO and the 2DVID conditions, meaning that the field-laboratory differences might be more apparent when measuring loud stimuli and undetectable for softer sounds. Clinical evaluations should pay special attention to these differences, as loud sounds are the ones that usually cause loudness discomfort.

Even though we found significant differences between the field and the laboratory ratings for one condition, these differences were below one loudness category / response alternative on average. Bigger and significant differences were expected, at least between the laboratory conditions [6]. One possible reason is that the participants had already seen and heard the driving actions in the field and could relate to them in the laboratory. Similarly, the participants did the laboratory conditions consecutively, meaning that some of them experienced the audiovisual conditions before the audio-only condition and could relate to the previous experience. For example, loudness ratings can be affected by the color of the vehicle [20]. Some participants might have been able to remember the vehicles and could picture them in a certain color, thus reducing differences between audiovisual and audio-only conditions. The physical-correlate theory [19] explains this phenomenon: the judgment of a sensory input is based upon experience and its physical correlate. Therefore, the results for the audiovisual conditions should be seen in conjunction with the whole experiment and not alone.

Although the field-laboratory differences were small on average, these differences should be considered when measuring loudness perception and during hearing-aid fitting procedures. In the following paragraphs the limitations and challenges of comparing field and laboratory loudness perception are described.

These should be considered when interpreting the results of this experiment.

### 4.1 Limitations

Making an exact replica of a field situation in the laboratory is very challenging, if not impossible [3], and requires expensive equipment and expertise [13]. In this experiment, we tried to reproduce the field stimuli in the laboratory as accurately as possible using a setup that can easily be set up for comparison in other labs or clinics. This means that marked differences between laboratory and field were present.

The participants were sitting in different positions in the field experiment. They did not see and hear the same stimuli as the recording devices. By being in a different sitting position, the sound pressure levels and the spectral shape of the driving actions changed. We tried to minimize this factor in the experimental design by doing the measurements in four different sessions, in order to have fewer participants per session and to have them sitting closer to the middle position and the recording devices. Nevertheless, we still had to remove about one third of the collected loudness ratings.

The driving actions were repeated eight times in the field and only one of those repetitions was used in the laboratory. Therefore, most participants did not experience the driving actions in the same way, as they were only present for two of those eight repetitions in the field. Nevertheless, the repetition of the driving actions was quite accurate in terms of sound pressure levels and the test-retest reliability of the ratings of the NH participants was high [7], therefore the effect on the ratings could be considered minimal.

The acoustic experience in the laboratory was not the same as in the field. In the laboratory, the sound was coming from one or two visible loudspeakers and although the room was acoustically treated, it was not fully anechoic. Acoustic reflections, room modes, and distance to the loudspeakers [14] could have affected the loudness ratings and added variability to the field-equivalent sound pressure levels. We wanted the design of our laboratory experiment to be closer to a clinical test than an exact reconstruction of the field experiment. Therefore, we did not provide any acoustic context in the laboratory: in the field experiment, the participants heard the vehicles when they were getting ready for each driving action and there was background noise between driving actions. They could expect a certain loudness, which did not happen in the laboratory.

### 4.2 Categorical Loudness Scaling

In our experiment we did not follow some of the standard procedures of categorical loudness scaling described by [10]. For example, the whole audible range should be presented (from not heard to too loud) and each signal should be presented at five different sound pressure levels. In our experiment, the quietest sound level was way above the hearing level (>65 dB SPL) and each driving action was presented at the same level for each laboratory condition. Nevertheless, we found that the categorical loudness scale to be the most convenient for our experiment.

The standard procedure calculates the average of the sound levels that belong to a loudness category. In our case, we calculated the average of the loudness categories for a condition once these were transformed to a numerical scale, to be able to compare between conditions. We assumed that the categorical units have a linear relationship with dB SPL and the loudness categories are equidistant in our experiment. The loudness function, i.e., the relationship between loudness categories and sound pressure levels, of narrow-band noise signals has been fitted in previous work using two straight lines [16]. For binaural broadband noise signals, the loudness function tends to be a single straight line [17]. Therefore, the linear relationship between loudness categories and sound pressure levels can be justified. Additionally, previous research has averaged categorical loudness ratings [18].

### 4.3 Future work

Future work should test the laboratory audiovisual conditions with participants that were not in the field, to avoid the physical-correlate effect. The hypothesis is that the rating differences between the audio-only and the audiovisual conditions will become significant, as in previous work [5]. Another possible experiment would be to let the participants adjust the volume/gain level of the stimuli. The hypothesis is that the chosen levels would be lower for the audio-only condition in comparison to the audiovisual conditions and to the levels recorded in the field.

A further improvement to this study design would be to add other urban vehicles such as electrical scooters. Such quieter vehicles would give references to the lower levels of the loudness scale and thus increase its validity, as in this study all the vehicles and driving actions were above 65 dB SPL and the quieter categories of the loudness scale were never used.

### ACKNOWLEDGMENTS

This work received funding from the EU's H2020 research and innovation program under the MSCA GA 675324 (ENRICH), from the Deutsche Forschungsgemeinschaft (DFG, Cluster of Excellence EXC 1077/1 "Hearing4all", and SFB1330 Projects B1 and C4). Thanks to Julia Schütze, Anja Kreuteler for helping conduct the experiments, and to Melanie Krüger for contacting the participants. Special thanks to the personnel of the old military facility.